\documentstyle[11pt]{article}
\newcommand{\be}[1]{\begin{equation}\label{#1}}                     
\newcommand{\ba}[1]{\begin{eqnarray}\label{#1}}                     
\newcommand{\ee}{\end{equation}}                                    
\newcommand{\ea}{\end{eqnarray}}                                    
\newcommand{\non}{\nonumber\\}                                      
\newcommand{\five}{\vspace{-0.5em}}                                 
\newcommand{\one}{\vspace{-0.1em}}                                  
\newcommand{\dis}{\displaystyle}                                    

\newcommand{\TR}{\mathop{\rm Tr}}
\newcommand{\Tr}{\mathop{\rm tr}}
\newcommand{\mm}{\mu} 
\newcommand{\lm}{\lambda-\mm}                                       
\newcommand{\freop}{\Bigl(\tilde I+\widetilde V\Bigr)} 
\newcommand{\freoplong}{\Bigl(\tilde I+\widetilde V
            -\frac{\alpha}{2\pi}\widetilde Y\Bigr)} 
\newcommand{\invop}{\Bigl(\tilde I-\widetilde R\Bigr)} 
\newcommand{\EP}[1]{E_+(#1|u)} 
\newcommand{\EL}{\langle E^L(\lambda)|} 
\newcommand{\ER}{|E^R(\mu)\rangle} 
\newcommand{\ELm}{\langle E^L(\mu)|}                                
\newcommand{\ERl}{|E^R(\lambda)\rangle}                             
\newcommand{\ERls}{E^R(\lambda)\rangle}                             
\newcommand{\ERs}{E^R(\mu)\rangle}                                  
\newcommand{\El}[1]{E_{#1}^L(\lambda|u)}                            
\newcommand{\Elv}[1]{E_{#1}^L(\lambda|v)}                           
\newcommand{\Er}[1]{E_{#1}^R(\mu|u)}                                
\newcommand{\Erl}[1]{E_{#1}^R(\lambda|u)}                                
\newcommand{\FL}{\langle F^L(\lambda)|}                             
\newcommand{\FR}{|F^R(\mu)\rangle}                                  
\newcommand{\FLm}{\langle F^L(\mu)|}                                
\newcommand{\FRl}{|F^R(\lambda)\rangle}                             
\newcommand{\FRls}{F^R(\lambda)\rangle}                             
\newcommand{\FRs}{F^R(\mu)\rangle}                                  
\newcommand{\Fl}[1]{F_{#1}^L(\lambda|u)}                            
                           
\newcommand{\Fr}[1]{F_{#1}^R(\mu|u)}                                
\newcommand{\Frl}[1]{F_{#1}^R(\lambda|u)}

\newcommand{\EM}[1]{E_-(#1|u)}                                      
\newcommand{\EMV}[1]{E_-(#1|v)}

\newcommand{\tV}{\widetilde V}                          
\newcommand{\tY}{\widetilde Y}

\newcommand{\hs}{\hat\sigma}                                        
\newcommand{\hb}{\hat b}                                        
\newcommand{\hB}{\widehat B}                                        
\newcommand{\hC}{\widehat C}                                        
\newcommand{\hD}{\widehat D}                                        
\newcommand{\hg}{\hat g}

\newcommand{\ketd}{|0)} 
\newcommand{\brad}{(0|}                                        
\newcommand{\Eq}[1]{(\ref{#1})}                                     
\newcommand{\stint}{\int\limits_{-\infty}^\infty}                   
\newcommand{\qint}{\int\limits_{-\infty}^\infty}

\newlength{\minitwocolumn}
\catcode`\@=11
\def\relaxnext@{\let\next\relax}
\def\eq#1\endeq{\begin{eqnarray}#1\end{eqnarray}}
\def\eqn#1\endeqn{\begin{eqnarray*}#1\end{eqnarray*}}

\makeatletter
\@addtoreset{equation}{section}
\makeatother

\begin{document}
%%%%%%%%%%%%%%%%%%%%%%%%%%%%%%%%%%%%%%%%%%%%%%%%%%%%%%%%%%
\begin{flushright}
\end{flushright}
\vspace{24pt}
\begin{center}
\begin{Large}
{\bf Completely Integrable Equation for the 

Quantum Correlation Function of Nonlinear Schr\"odinger Eqaution.}
\end{Large}

\vspace{36pt}

T.~Kojima\raisebox{2mm}{{\scriptsize a}$\star$},
V.~E.~Korepin\raisebox{2mm}{{\scriptsize b}}

and N.~A.~Slavnov\raisebox{2mm}{{\scriptsize c}}

\vspace{6pt}

~\raisebox{2mm}{{\scriptsize a}}
{\it Research Institute for Mathematical Sciences,
     Kyoto University, Kyoto 606, Japan}

~\raisebox{2mm}{{\scriptsize b}}
{\it Institute for Theoretical Physics, State University of 
New York at Stony Brook,
Stony Brook, NY 11794-3840, U. S. A.}

~\raisebox{2mm}{{\scriptsize c}}
{\it Steklov Mathematical Institute,
Gubkina 8, Moscow 117966, Russia.}
\vspace{72pt}

\underline{Abstract}
\end{center} 
Correlation functions of exactly solvable models can be described by 
differential equations \cite{BMW}. In this paper we show that for 
non free fermionic case differential equations should be replaced by 
integro-differential equations.  We derive an integro-differential 
equation, which describes time and temperature dependent correlation 
function $\langle\psi(0,0)\psi^\dagger(x,t)\rangle_T$ of penetrable 
Bose gas. The integro-differential equation turns out be the continuum
generalization of classical nonlinear Schr\"odinger equation.



Differential eqautions; Lax Pair; Quantum Inverse Scattering Method

\vspace{24pt}

\vfill
\hrule

\vskip 3mm
\begin{footnotesize}
\noindent
\raisebox{2mm}{$a$}
kojima@kurims.kyoto-u.ac.jp,~
\raisebox{2mm}{$b$}
korepin@insti.physics.sunysb.edu,~
\raisebox{2mm}{$c$}
nslavnov@class.mi.ras.ru\\
\noindent\raisebox{2mm}{$\star$}
Research Fellow of the Japan Society
for the Promotion of Science.
\end{footnotesize}
\newpage
%%%%%%%%%%%%%%%%%%%%%%%%%%%%%%%%%%%%%%%%%%%%%%
\section{Introduction}
%%%%%%%%%%%%%%%%%%%%%%%%%%%%%%%%%%%%%%%%%%%%%%%%%%%%%%%%%%%
We consider  exactly solvable models of statistical mechanics in one space
and one time dimension. The Quantum Inverse Scattering Method
and Algebraic Bethe Ansatz are effective methods for a description of the
spectrum of these models. Our aim is the evaluation of correlation functions of
exactly solvable models. Our approach is based on the determinant representation for correlation functions. It consists of a few steps: first the correlation
function is represented as a determinant of a Fredholm integral operator,
second  the Fredholm integral operator is described by a 
classical completely
integrable equation, third the classical completely integrable equation
is solved by means of the Riemann--Hilbert problem. This permits us to evaluate
the long distance and large time asymptotics of the correlation function.
The method is described in \cite{K.B.I.}.
The most interesting correlation functions
are time dependent correlation functions. The determinant
representation for  time and temperature dependent correlation functions
of quantum nonlinear Schr\"odinger equation was obtained in \cite{KKS}.
 In this paper we describe
the correlation function by means of completely integrable 
integro-differential equation. In the forthcoming publication we 
shall formulate  the  Riemann--Hilbert problem for this equation and 
evaluate long distance asymptotic.

The  quantum nonlinear Schr\"odinger equation  can be
described in terms of canonical Bose fields
$\psi(x,t), \psi^{\dagger}(x,t)~(x \in {\bf R})$ obeying
\be{compsi}
[\psi(x,t), \psi^{\dagger}(y,t)]=\delta(x-y).
\ee
The Hamiltonian and momentum of the model are
\ba{Hamilton}
{H}&=&{\dis\int \,dx
\left({\partial_x}\psi^{\dagger}(x)
{\partial_x} \psi(x)+
c\psi^{\dagger}(x)\psi^{\dagger}(x)\psi(x)\psi(x)
-h \psi^{\dagger}(x)\psi(x)\right),}\\
\one&\one&\one\non
\label{momentum}
P&=&-i\int\,dx\psi^\dagger(x)\partial_x\psi(x).
\ea
Here $0<c\le\infty$ is the coupling constant and $h>0$
is the chemical potential.
The spectrum of the model was first described by E.~H.~Lieb and
 W.~Liniger \cite{L.L.}, \cite{L}.
The Lax representation for the corresponding classical equation of 
 motion
\begin{eqnarray}
i\frac{\partial}{\partial t} \psi=
[\psi, H]=
-\frac{\partial^2}{{\partial x}^2} \psi
+2c \psi^{\dagger}\psi \psi
-h \psi,
\end{eqnarray}
was found by V.~E.~Zakharov and A.~B.~Shabat \cite{ZS}. 
The Quantum Inverse Scattering Method for the model
was formulated by L.~D.~Faddeev and E.~K.~Sklyanin  \cite{FS}.
The quantum nonlinear Schr\"odinger equation is equivalent to the Bose gas
with delta-function interaction. In the sector with $N$ particles the
Hamiltonian of Bose gas is given by
\begin{eqnarray}
{\cal H}_N=-\sum_{j=1}^N
\frac{\partial ^2}{\partial z_j^2}
+2c \sum_{1 \leq j < k \leq N}
\delta(z_k-z_j)-Nh.
\end{eqnarray}
In this paper we shall consider  the thermodynamic of the model. 
The  partition function and  the free energy of the model are 
defined by 
\be{partfunct} Z=\Tr e^{-\frac{H}{T}} = e^{-\frac{F}{T}}.  
\ee 
The  free energy $F$ can be expressed in terms of Yang-Yang 
equation \cite{YY} 
\ba{YangYang} 
\varepsilon(\lambda)&=&{\dis\lambda^2-h-
\frac{T}{2\pi}\stint\frac{2c}{c^2
+(\lambda-\mu)^2}\ln\left(1+e^{-\frac
{\varepsilon(\mu)}T}\right)\,d\mu,}\\
\one&\one&\one\non
\label{freeenergy}
F&=&{\dis
-\frac{T}{2\pi}\stint
\ln\left(1+e^{-\frac
{\varepsilon(\mu)}T}\right)\,d\mu.}
\ea
The correlation function, which we shall study in this paper, is
defined by
\be{tempcorrel}
\langle\psi(0,0)\psi^\dagger(x,t)\rangle_T=
\frac{\Tr\left( e^{-\frac HT}\psi(0,0)\psi^\dagger(x,t)\right)
}
{\Tr e^{-\frac HT}}.
\ee

In the previous paper \cite{KKS} we obtained the determinant representation 
for this correlation function. In this paper we shall derive completely 
integrable equations, starting from  the determinant representation. The 
plan of this paper is the following. In Section 2 we shall remind the
reader the determinant representation and definition of dual fields.
In Section 3 we introduce new Hilbert space and rewrite the kernel of 
the integral operator in the canonical form. In Section 4 we define the
resolvent of the integral operator.  Section 5 is devoted to the 
construction of the Lax representation. In Section 6 we find the 
logarithmic derivatives of the Fredholm determinant and obtain 
 the completely integrable equation describing the correlation function. 
In Section 7 we summarize the main results. In Appendix A we find some
identities for potentials. We give the treatment of  the quantum 
nonlinear Schr\"odinger equation as a continuum generalization of 
the classical equation in Appendix B. Appendix C is devoted to the 
free fermion limit.

%%%%%%%%%%%%%%%%%%%%%%%%%%%%%%%%%%%%%%%%%%%%%%%%%%%%%%%%%%
\section{Determinant representation for the correlation function} 
%%%%%%%%%%%%%%%%%%%%%%%%%%%%%%%%%%%%%%%%%%%%%%%%%%%%%%%%%%%
Our starting point is the determinant representation for the 
temperature correlation function of local fields obtained in
\cite{KKS}
\newpage
 \ba{corr1}&&{\dis \langle 
\psi(0,0)\psi^\dagger(x,t)\rangle_T= e^{-iht}\brad \left( 
G(x,t)+\frac{\partial}{\partial \alpha}\right)}\non 
&&\hspace{3cm}{\dis\times 
\left.\frac{\det 
\left(\tilde I+\tV-\frac{\alpha}{2\pi} 
\widetilde Y\right)}{\det\left(\tilde 
I-\frac{1}{2\pi}\widetilde K_T\right)}\ketd\right|_{\alpha=0}.}
\ea 
Let us explain our notations.

We begin by the numerator in the r.h.s. of \Eq{corr1}, is the Fredholm 
determinant of the integral operator
$\tilde I+\tV-\frac{\alpha}{2\pi}\widetilde Y$ (here 
$\tilde I$ is identical operator: $I(\lambda,\mu)=
\delta(\lambda-\mu)$). This operator acts on the real axis. The left 
and right actions on some trial function $f$ are given by
\be{intopact}
\begin{array}{c}
{\dis\freoplong\circ f(\mu)=f(\lambda)+\qint 
\left(V(\lambda,\mu)-\frac{\alpha}{2\pi}Y(\lambda,\mu)\right)
f(\mu)\,d\mu,}\\
\five\\
{\dis f(\lambda)\circ\freoplong=f(\mu)+\qint 
f(\lambda)\left(V(\lambda,\mu)
-\frac{\alpha}{2\pi}Y(\lambda,\mu)\right)
\,d\lambda.}
\end{array}
\ee
Here and hereafter we denote  by the symbol ``$\circ$" the action of 
integral operators on functions.

The kernels of operators $\tV$ and $\tY$  can be written in terms of 
auxiliary quantum operators --- dual fields, acting in an 
auxiliary Fock space. One can find the detailed definition and 
properties of dual fields in Section 5 and Appendix C of \cite{KKS}. 
Here we repeat them in brief.

Consider an auxiliary Fock space having vacuum vector $\ketd$ and 
dual vector $\brad$. Three dual fields $\psi(\lambda)$, 
$\phi_{D_1}(\lambda)$ and $\phi_{A_2}(\lambda)$ acting in this space 
are defined as
\be{dualfields}
\begin{array}{rcl}
\phi_{A_2}(\lambda)&=&q_{A_2}(\lambda)+p_{D_2}(\lambda),\\
\phi_{D_1}(\lambda)&=&q_{D_1}(\lambda)+p_{A_1}(\lambda),\\
\psi(\lambda)&=&q_\psi(\lambda)+p_\psi(\lambda).
\end{array}
\ee
Here $p(\lambda)$ are annihilation parts of dual fields:
$p(\lambda)\ketd=0$;
$q(\lambda)$ are creation parts of dual fields: $\brad q(\lambda)=0$.
Thus, any dual field is the sum of annihilation and creation parts.
Nonzero commutation relations are (see \cite{KKS})
\be{commutators}
\begin{array}{rcl}
{}[p_{A_1}(\lambda),q_\psi(\mu)]&=&\ln h(\mu,\lambda),\\
{}[p_{D_2}(\lambda),q_\psi(\mu)]&=&\ln h(\lambda,\mu),\\
{}[p_\psi(\lambda),q_{A_2}(\mu)]&=&\ln h(\mu,\lambda),
\qquad\qquad\mbox{where}\qquad
 {\dis h(\lambda,\mu)=\frac{ic}{\lambda-\mu+ic}}\\
{}[p_\psi(\lambda),q_{D_1}(\mu)]&=&\ln h(\lambda,\mu),\\
{}[p_\psi(\lambda),q_\psi(\mu)]&=&\ln [h(\lambda,\mu)h(\mu,\lambda)].
\end{array}
\ee
Recall that $c$ is the coupling constant in \Eq{Hamilton}.
It follows immediately from \Eq{commutators} that the dual fields 
belong to an Abelian sub-algebra 
\be{Abel} [\psi(\lambda),\psi(\mu)]= 
[\psi(\lambda),\phi_a(\mu)]=
[\phi_b(\lambda),\phi_a(\mu)]=0,
\ee
where $a,b=A_2,D_1$. This property, in fact, permits us to treat
the dual fields as some $c$-number functions. 

Let us define function $Z(\lambda,\mu)$:
\be{Z}
Z(\lambda,\mm)=\frac{e^{-\phi_{D_1}(\lambda)}}{h(\mm,\lambda)}+
\frac{e^{-\phi_{A_2}(\lambda)}}{h(\lambda,\mm)}.
\ee
The kernel $\tV$ is equal to
 \ba{mainkernel}&&{\dis 
V(\lambda,\mu)=\frac{
e^{\frac{1}{2}(\phi_{D_1}(\lambda)+\phi_{A_2}(\lambda))}
e^{\frac{1}{2}(\phi_{D_1}(\mu)+\phi_{A_2}(\mu))}
\sqrt{\theta(\lambda)}\sqrt{\theta(\mu)}}
{4\pi^2(\lambda-\mu)} }\non
\five\non
&&{\dis
\times\stint \,\frac{du}{Z(u,u)}
\left(\frac{e^{-\phi_{D_1}(u)}}{u-\lambda-i0}+
\frac{e^{-\phi_{A_2}(u)}}{u-\lambda+i0}-
\frac{e^{-\phi_{D_1}(u)}}{u-\mu-i0}- 
\frac{e^{-\phi_{A_2}(u)}}{u-\mu+i0}\right)
}\non 
\five\non
&&\hskip3cm{\dis \times
e^{\psi(u)+\tau(u)}
e^{-\frac{1}{2}(\psi(\lambda)+\tau(\lambda)+\psi(\mu)+\tau(\mu))}
Z(u,\lambda)Z(u,\mu).}
\ea
Integral operator $\tY$ is one-dimensional projector
\be{tY}
Y(\lambda,\mu)=P(\lambda)P(\mu),
\ee
where
\ba{project1}&&{\dis
 P(\mm)=\frac{e^{\frac{1}{2}(\phi_{D_1}(\mu)+\phi_{A_2}(\mu))}
\sqrt{\theta(\mu)}}{2\pi}
\stint \,\frac{du}{Z(u,u)}
\left(\frac{e^{-\phi_{D_1}(u)}}{u-\mm-i0}+ 
\frac{e^{-\phi_{A_2}(u)}}{u-\mm+i0}\right)}\non
\five\non
&&{\dis\hskip3.5cm
\times e^{\psi(u)+\tau(u)}
e^{-\frac{1}{2}(\psi(\mu)+\tau(\mu))} Z(u,\mm).}
\ea
Here functions $\theta(\lambda)$ and $\tau(\lambda)$ are equal to
\ba{theta}
\theta(\lambda)&=&
{\dis\left(1+\exp\left[\frac{\varepsilon(\lambda)}T\right]\right)^{-1}},\\
\one&\one&\one\non
\label{tau}
\tau(\lambda)&=&it\lambda^2-ix\lambda.
\ea
The Fermi weight $\theta(\lambda)$ defines the dependence of 
 the correlation function on temperature $T$. The energy of 
one-particle excitation $\varepsilon(\lambda)$ is given in 
\Eq{YangYang}. The  function $\tau(\lambda)$ depends also on the 
distance $x$ and the time $t$. All other functions entering  
expressions for $V(\lambda,\mu)$ and $P(\mu)$ do not depend on 
$x$ and $t$.

We would like to draw reader's attention to the fact that 
formul\ae~\Eq{mainkernel} and \Eq{project1} slightly differ from 
formul\ae~(6.25) and (6.26) of \cite{KKS}. It is explained in the
Appendixes C and D of \cite{KKS} how one can reduce
formul\ae~(6.25) and (6.26) to formul\ae~
\Eq{mainkernel} and \Eq{project1}. Thus,  the integral
operator $\tilde I+\tV-\frac{\alpha}{2\pi}\tY$ is explained.

The operator $\tilde I-\frac{1}{2\pi}\widetilde K_T$ also acts on the
whole real axis. Its kernel is given by
\be{KT}
K_T(\lambda,\mu)=\left(\frac{2c}{c^2+
(\lambda-\mu)^2}\right)
\sqrt{\theta(\lambda)}\sqrt{\theta(\mu)}.
\ee
Finally  the function $G(x,t)$ in the r.h.s. of \Eq{corr1} is equal to
\be{G1}
G(x,t)=\frac{1}{2\pi}\stint e^{\psi(v)+\tau(v)}\,dv.
\ee

Thus, we have described the r.h.s. of \Eq{corr1}. The temperature 
correlation function of local fields is proportional to the
vacuum expectation in  the auxiliary Fock space of the Fredholm 
determinant of the integral operator. The  auxiliary quantum 
operators---dual fields---enters  the kernels $\tV$ and $\tY$. 
Due to the property \Eq{Abel} the Fredholm determinant is well 
defined.  Our aim now is a description of the correlation function in 
terms of solutions of classical completely integrable equations.

%%%%%%%%%%%%%%%%%%%%%%%%%%%%%%%%%%%%%%%%%%%%%%%%%%%%%%%%%% 
\section{Vectors and operators of new Hilbert space}
%%%%%%%%%%%%%%%%%%%%%%%%%%%%%%%%%%%%%%%%%%%%%%%%%%%%%%%%%%

Introducing functions $E_\pm$:
\ba{eplus}
&&{\dis
\EP\lambda=\frac{1}{2\pi}\frac{Z(u,\lambda)}{Z(u,u)}
\left(\frac{e^{-\phi_{A_2}(u)}}{u-\lambda+i0}
+\frac{e^{-\phi_{D_1}(u)}}{u-\lambda-i0}\right)\sqrt{\theta(\lambda)}}\non
\one &\one &\one \non
&&\hskip3cm{\dis \times
e^{\psi(u)+\tau(u)+\frac12(
\phi_{D_1}(\lambda)+\phi_{A_2}(\lambda)-\psi(\lambda)-\tau(\lambda))},}
\ea
\be{eminus}
\EM\lambda=\frac{1}{2\pi}Z(u,\lambda)
e^{\frac12(\phi_{D_1}(\lambda)+\phi_{A_2}(\lambda)
-\psi(\lambda)-\tau(\lambda))}\sqrt{\theta(\lambda)}.
\ee
The  functions $E_\pm$ depend also on  the distance $x$, 
the time $t$,  the temperature $T$ and  the chemical 
potential $h$, but this dependence as a rule is suppressed in the 
notation. One can rewrite expressions \Eq{mainkernel} and 
\Eq{project1} for $V(\lambda,\mu)$ and $P(\mu)$ in terms of these 
functions

\be{kernelV} V(\lambda,\mm)=\frac{1}{\lm}\stint\,du 
(\EP\lambda\EM\mm-\EM\lambda\EP\mm),
\ee
\be{kernelP}
Y(\lambda,\mu)= P(\lambda)P(\mm)=\stint\,dudv\EP\lambda 
E_+(\mm|v).  
\ee 
Using obvious equalities
$$
\partial_xe^{\tau(\lambda)}
=-i\lambda e^{\tau(\lambda)};\qquad
\partial_te^{\tau(\lambda)}
=i\lambda^2 e^{\tau(\lambda)},
$$
we arrive at
\be{derivativeEPx}
\begin{array}{rcl}
\partial_x\EP\lambda&=&-\frac{i\lambda}{2}\EP\lambda-
i\stint\,dv g(u,v)\EMV\lambda,\\
\one&\one&\one\\
\partial_x\EM\lambda&=&\frac{i\lambda}{2}\EM\lambda,
\end{array}
\ee
\be{derivativeEPt}
\begin{array}{rcl}
\partial_t\EP\lambda&=&\frac{i\lambda^2}{2}\EP\lambda
+\stint\,dv(i\lambda g(u,v)-\partial_x g(u,v))\EMV\lambda,\\
\one&\one&\one\\
\partial_t\EM\lambda&=&-\frac{i\lambda^2}{2}\EM\lambda.
\end{array}
\ee
Here
\be{kernelg}
g(u,v)=\delta(u-v)e^{\psi(v)+\tau(v)}.
\ee
The relations \Eq{derivativeEPx} and \Eq{derivativeEPt} are 
important for a description of the correlation function in terms of 
solutions of completely integrable equations. In order to do this it 
is necessary to investigate properties of the integral operator 
$\tilde I+\tV$. Later  on we shall show how one can take into 
consideration the contribution of the projector $\tY$.

In order to derive differential equations for the correlation function
it is convenient to treat functions $E_\pm$ as components of vectors 
of some Hilbert space ${\cal H}$, for example, 
rigged $L_2(-\infty,\infty)\otimes R_2$ \cite{GV}.  Let us 
introduce the bra-vector  $\EL$ and the ket-vector $\ERl$ belong to 
the Hilbert space ${\cal H}$.  Both of these vectors have two 
discrete components (corresponding to the space $R_2$), which we 
shall denote with indices $1$ and $2$.  In turn any discrete 
component has continuous ``index" (corresponding to the rigged 
$L_2(-\infty,\infty)$ space) which we shall denote as ``$u$" (or ``$v$", 
``$w$" etc.).  So 
$$ \EL=\Bigl(\El1,\El2\Bigr);\qquad\qquad 
\ER=\left(\begin{array}{c} \Er1\\ {}\\ \Er2 \end{array}\right).  
$$ 
The definition of  the scalar product is standard :  
\be{scalprod} 
\EL\ERs=\stint\,du\left(\El1\Er1+\El2\Er2\right).
\ee
On the contrary the products of type $\ER\EL$ as usual are operators in 
Hilbert space ${\cal H}$. We shall consider such operators below.

Let us identify
\ba{identification}
\Er1=\EP\mu,&\qquad\qquad&\El1=-\EM\lambda,\non
\vspace{-0.1em}&\vspace{-0.1em}&\vspace{-0.1em}\\
\Er2=\EM\mu,&\qquad\qquad&\El2=\EP\lambda.\nonumber
\ea
Then one can rewrite  the kernel of the integral operator $V$ as
\be{kernel}
V(\lambda,\mu)=\frac{\EL\ERs}{\lm}.
\ee
Due to \Eq{identification} we have
\be{orthogonality}
\EL\ERls=0
\ee
and hence the kernel $V(\lambda,\mu)$ is not singular in the point
$\lambda=\mu$.

The representation \Eq{kernel}  is the canonical form 
of the kernels of  the completely integrable integral operators.  In all 
examples related to correlation functions, the kernels of integral 
operators can be presented in the form \Eq{kernel}. The different
realizations of space ${\cal H}$ correspond to the concrete correlation
functions. For example, in  the free   
fermion situation (the coupling constant
$c$ goes to infinity) ${\cal H}=R_n$, where n is the number of fields
\cite{IIK1}, \cite{IIK2}, \cite{IIKS}, \cite{K.S.1}, \cite{S}. For 
equal-time correlation functions of penetrable bosons the 
representations of type \Eq{kernel} were constructed with ${\cal 
H}=L_2(0,\infty)\otimes R_{2n}$  in \cite{IIKS}, \cite{K}, \cite{KS} 
(see also Section XIV of \cite{K.B.I.}). In the present paper we 
shall follow the method developed in the papers enumerated above.

Operators acting in the space ${\cal H}$ are defined in the standard way.  
They have  discrete and continuous indices:  $\widehat 
A=A_{jk}(u,v),\qquad j,k=1,2;\quad-\infty<u,v<\infty$. We shall 
denote these operators with the sign ``hat" in order to distinguish them 
from integral operators which we have denoted with the sign ``tilde".  
Action on vectors is given by 
\ba{action} && \widehat 
A\ER=\sum_{k=1}^{2}\stint A_{jk}(u,v)E_k^R(\mu|v)\,dv,\non % 
&&\vspace{-0.5em}\\
&&
\EL\widehat A=\sum_{j=1}^{2}\stint
\El{j}A_{jk}(u,v)\,du.\nonumber
\ea
On the contrary
the integral operators of type $\tilde I+\tV$ appear to be scalars 
relatively to the space ${\cal H}$, for example 
$$ 
\freop\circ\ER=\left( \begin{array}{c} {\dis \Erl1+\qint 
V(\lambda,\mu)\Er1\,d\mu}\\ \five\\ {\dis \Erl2+\qint 
V(\lambda,\mu)\Er2\,d\mu} \end{array} \right).  
$$
The product of operators of type $\hat A$ is
$$
\widehat A^{(1)}\widehat A^{(2)} =
\sum_{l=1}^{2}\stint A_{jl}^{(1)}(u,w)A_{lk}^{(2)}(w,v)\,dw.
$$
The trace of operator is defined as usual
$$
\Tr\widehat A=\stint \,du\biggl(
A_{11}(u,u)+A_{22}(u,u)\biggr).
$$
In particular for operators of type $|\dots\rangle\langle\dots|$ we 
have 
$$ 
\Tr|\dots\rangle\langle\dots|=\langle\dots|\dots\rangle.  
$$

Using  the operator  notations, one can rewrite the relations 
\Eq{derivativeEPx}, \Eq{derivativeEPt} in the form
of linear partial differential equations : 
\be{derivativexE} 
\partial_x\ERl=\widehat L(\lambda)\ERl,\qquad
\partial_x\EL=-\EL\widehat L(\lambda),
\ee
\be{derivativetE}
\partial_t\ERl=\hat M(\lambda)\ERl,\qquad
\partial_t\EL=-\EL\hat M(\lambda),
\ee
where
\be{L}
\widehat L(\lambda)=\lambda\hs+[\hg,\hs],
\ee
\be{M}
\hat M(\lambda)=-\lambda\widehat L(\lambda)+\partial_x\hg,
\ee
and  the operators $\hs$ and $\hg$ are equal to
\be{sigma}
\hs=-\frac{i}{2}\sigma_3\delta(u-v)
=-\frac{i}{2}\left(\begin{array}{rr}
1&0\\0&-1
\end{array}\right)\delta(u-v),
\ee
\be{q}
\hg=-\sigma_+g(u,v)
=\left(\begin{array}{rr}
0&-1\\0&0
\end{array}\right)g(u,v).
\ee
Later we shall use  the relations \Eq{derivativexE} and \Eq{derivativetE} 
in order to construct the nontrivial Lax representation.
%%%%%%%%%%%%%%%%%%%%%%%%%%%%%%%%%%%%%%%%%%%%%%%%%%%%%%%
\section{Vectors $\FL,\quad \FRl$ and resolvent}
%%%%%%%%%%%%%%%%%%%%%%%%%%%%%%%%%%%%%%%%%%%%%%%%%%%%%%%
Let us introduce  the vectors 
$\FL$ and $\FRl$ belong to the same  space 
${\cal H}$
\be{introF} 
\FL=\Bigl(\Fl1,\Fl2\Bigr);\qquad\qquad 
\FR=\left(\begin{array}{c}
\Fr1\\
{}\\ 
\Fr2
\end{array}\right),
\ee
defining them as solutions of  the integral equations
\ba{intequation1}
&&\freop\circ\FLm=\EL,\\
&&\five\non
&&\FRl\circ\freop=\ER.
\label{intequation2}
\ea
More preciously these formul\ae\ mean
$$
\begin{array}{l}{\dis
F_j^L(\lambda)+\qint 
V(\lambda,\mu)F_j^L(\mu)\,d\mu=E_j^L(\lambda),}\\ 
\five\\ 
{\dis 
F_j^R(\mu)+\qint F_j^R(\lambda)
 V(\lambda,\mu)\,d\lambda=E_j^R(\mu).}
\end{array}
$$

Define the resolvent of the operator $\tilde I-\tV$ as
\be{inverse}
\freop\circ\Bigl(\tilde I-\widetilde R\Bigr)=\tilde I.
\ee
Obviously
\ba{intequation3}
&&\invop\circ\ELm=\FL,\\
&&\five\non
&&\ERl\circ\invop=\FR.
\label{intequation4}
\ea
Let us find the kernel of the resolvent. One can rewrite \Eq{inverse} 
as follows
$$
V(\lambda,\mu)-\qint
V(\lambda,\nu)R(\nu,\mu)\,d\nu=R(\lambda,\mu).
$$
Multiplying both sides of the last equality by $\lm$ we get
$$
(\lm)V(\lambda,\mu)-\qint
V(\lambda,\nu)(\lambda-\nu+\nu-\mu)R(\nu,\mu)\,d\nu
=(\lm)R(\lambda,\mu),
$$
or
$$
\EL\ERs-\qint\EL E^R(\nu)\rangle R(\nu,\mu)\,d\nu=
\freop\circ(\nu-\mu)R(\nu,\mu),
$$
or due to \Eq{intequation4}
$$
\EL\FRs=\freop\circ(\nu-\mu)R(\nu,\mu).
$$
Making $\tilde{I}-\widetilde{R}$ act on this equation from the left, we get
$$
\FL\FRs=(\lm)R(\lambda,\mu).
$$
Therefore we have come to the following expression for the resolvent
kernel:
\be{invoperator}
R(\lambda,\mu)=\frac{\FL\FRs}{\lm}.
\ee
It is worth mentioning that this method of calculation of resolvent
is a direct generalization of the method described in the Section
XIV.1 of \cite{K.B.I.}.

In the next Section we shall need  the operator  $\hB$ 
(potential), defined as 
\be{B1} \hB=\qint\FRl\EL\,d\lambda.  
\ee 
Obviously 
\be{B2} 
\hB=\FRl\circ\freop\circ\FLm=
\qint\ER\FLm\,d\mu.
\ee
The  components of this operator are
\be{componentB}
B_{jk}(u,v)=\qint\Frl{j}\Elv{k}\,d\lambda,
\ee
so
\be{matrixB}
\hB=\left(
\begin{array}{cc}
B_{11}(u,v) & B_{12}(u,v)\\
B_{21}(u,v) & B_{22}(u,v)
\end{array}
\right).
\ee
The  operator $\hC$ defined in the similar way is also useful
\be{hC}
\hC=\qint\lambda\FRl\EL\,d\lambda.
\ee
The  components of the operator $\hC$ are
\be{componentC}                                          
C_{jk}(u,v)=\qint\lambda\Frl{j}\Elv{k}\,d\lambda,
\ee
so
\be{matrixC}
\hC=\left(
\begin{array}{cc}
C_{11}(u,v) & C_{12}(u,v)\\
C_{21}(u,v) & C_{22}(u,v)
\end{array}
\right).
\ee
%
%%%%%%%%%%%%%%%%%%%%%%%%%%%%%%%%%%%%%%%%%%%%%%%%%%%%%%%%%%
\section{The  Lax representation}
%%%%%%%%%%%%%%%%%%%%%%%%%%%%%%%%%%%%%%%%%%%%%%%%%%%%%%%%%%%%

In this Section we construct the Lax representation having nontrivial 
compatibility condition. Na\-me\-ly, we establish the following 
relations 
\be{derivativexFprim} \partial_x\FRl=\widehat{\cal 
L}(\lambda)\FRl, 
\ee 
\be{derivativetFprim} \partial_t\FRl=\hat{\cal 
M}(\lambda)\FRl, 
\ee 
which are analogous to the relations 
\Eq{derivativexE} and \Eq{derivativetE}. We shall prove that one can 
obtain  the operators $\widehat{\cal L}(\lambda)$ and $\hat{\cal 
M}(\lambda)$ using formul\ae~\Eq{L} and \Eq{M} with replacement $\hg$ 
by $\hg+\hB$.

Let us calculate  the derivative of $V(\lambda,\mu)$ with
respect to $x$ using formul\ae\ \Eq{derivativexE}
\ba{derivativexV}
\partial_x V(\lambda,\mu)&=&
-\frac{\EL(\widehat L(\lambda)-\widehat L(\mu))\ER}{\lm}\non
\one & \one & \one \non
&=&-\EL\hs\ER.
\ea
Thus, from \Eq{intequation2} we get
\be{intermediate1} 
\partial_x\FRl\circ\freop-\qint\FRl\EL\hs\ER\,d\lambda
=\widehat L(\mu)\ER,
\ee
or using  the definition of $\hB$
$$
\partial_x\FRl\circ\freop-\hB\cdot\hs\ER
=\widehat L(\mu)\ER.
$$
Making $\tilde I-\widetilde R$ on this equality the right, we get
$$
\partial_x\FRl-\hB\cdot\hs\FRl
=(\widehat L(\mu)-\widehat L(\lambda))
\ER\circ\invop+\widehat L(\lambda)\FRl.
$$
In the r.h.s. we have
$$
\begin{array}{rcl}
(\widehat L(\mu)-\widehat L(\lambda))\ER\circ\invop&=&
-\hs\qint\ER\FLm\FRls\,d\mu\\
\one&\one&\one\\
&=&-\hs\cdot\hB\FRl.
\end{array}
$$
Therefore
$$
\partial_x\FRl=\left(\widehat L(\lambda)+
[\hB,\hs]\right)\FRl,
$$
or
\be{derivativexF}
\partial_x\FRl=\widehat{\cal L}(\lambda)\FRl,
\ee
where
\be{calL}
\widehat{\cal L}(\lambda)=\lambda\hs+[\hb,\hs],
\ee
and
\be{b}
\hb=\hB+\hg.
\ee
In the same way one can obtain the similar formula for $\FL$
$$
\partial_x\FL=-\FL\widehat{\cal L}(\lambda).
$$

Now let us turn to the derivative of $\FRl$ with respect to $t$.
As before, we start with differentiation of the kernel $V(\lambda,\mu)$ 
using formul\ae\ \Eq{derivativetE}
\ba{derivativetV}
\partial_t V(\lambda,\mu)&=&
\frac{\EL(\lambda\widehat L(\lambda)-\mu\widehat L(\mu))\ER}{\lm}\non
\one & \one & \one \non
\hskip2cm&=&
\frac{\EL((\lambda-\mu)\widehat L(\lambda)-\mu(\widehat 
L(\mu)-\widehat L(\lambda))\ER}{\lm}\non 
\one & \one & \one \non
&=&
\EL\widehat L(\lambda)\ER+\mu\EL\hs\ER\non
\one & \one & \one \non
&=&
-\partial_x\EL\cdot\ER+\mu\EL\hs\ER.
\ea 
Thus, from \Eq{intequation2} we get
\ba{intermediate2}
&& {\dis 
\partial_t\FRl\circ\freop-\qint\FRl\partial_x\EL\cdot\ER\,d\lambda
}\non
\one  &\one &\one \non
&&{\dis
+\mu\qint\FRl\EL\hs\ER\,d\lambda
=[-\mu\widehat L(\mu)+\partial_x\hg]\ER.}
\ea
Comparing \Eq{intermediate2} with \Eq{intermediate1}, we see that
$$
\begin{array}{c}{\dis
\mu\qint\FRl\EL\hs\ER\,d\lambda+
\mu\widehat L(\mu)\ER}\\
\five\\
{\dis
\hskip6cm=\partial_x\FRl\circ\freop\cdot\mu.}
\end{array}
$$
Substituting this formula into \Eq{intermediate2} we find
\ba{intermediate3}&&{\dis 
\partial_t\FRl\circ\freop+\partial_x\FRl\circ\freop\cdot\mu}\non
\one &\one &\one  \non
&&{\dis
=\qint\FRl\partial_x\EL\cdot\ER\,d\lambda
+\partial_x\hg\ER.}
\ea
Acting on \Eq{intermediate3} by the resolvent from the right we have
\ba{intermediate4} &&{\dis
\partial_t\FRl+\partial_x\FRl\circ\freop\cdot\mu\circ\invop}\non
\one &\one & \one \non
&&{\dis
=\left(\qint\FR\partial_x\ELm\,d\mu\right)\cdot\FRl
+\partial_x\hg\FRl.}
\ea
The second term in the l.h.s. of \Eq{intermediate4} is equal to
$$
\begin{array}{c}{\dis
\partial_x\FRl\circ\freop\cdot\mu\circ\invop}\\
\five\\
{\dis
=\partial_x\FRl\circ\freop\cdot(\mu-\lambda+\lambda)\circ\invop}\\
\five\\
{\dis
=\lambda\partial_x\FRl-
\partial_x\FRl\circ\freop\circ\FLm\FRls}\\
\five\\
{\dis
=\lambda\partial_x\FRl-
\left(\qint\partial_x\FR\cdot\ELm\,d\mu\right)\FRl}.
\end{array}
$$
Therefore we arrive at
\ba{subcalM} {\dis
\partial_t\FRl}&=&{\dis
-\lambda\partial_x\FRl+
\left(\qint\partial_x\FR\cdot\ELm\,d\mu\right)\cdot\FRl}\non
\one &\one & \one \non
&+&{\dis
\left(\qint\FR\cdot\partial_x\ELm\,d\mu\right)\cdot\FRl
+\partial_x\hg\FRl}\non
\one & \one & \one \non
&=&
{\dis
-\lambda\partial_x\FRl+\partial_x(\hB+\hg)\FRl,}
\ea
or
\be{derivativetF}
\partial_t\FRl=\hat{\cal M}(\lambda)\FRl.
\ee
Here
\be{calM}
\hat{\cal M}(\lambda)=-\lambda\widehat{\cal L}(\lambda)+
\partial_x\hb,
\ee
or, using \Eq{calL} we can rewrite \Eq{calM} as
\be{calM1}
\hat{\cal M}(\lambda)=
-\lambda^2\hs-\lambda[\hb,\hs]+\partial_x\hb.
\ee

In the same way one can obtain the similar formula for $\FL$
$$
\partial_t\FL=-\FL\hat{\cal M}(\lambda).
$$
Thus, we have constructed the Lax representation.
%%%%%%%%%%%%%%%%%%%%%%%%%%%%%%%%%%%%%%%%%%%%%%%%%%%%%%%%
\section{The logarithmic derivatives of  the 
determinant and  the differential equations} 
%%%%%%%%%%%%%%%%%%%%%%%%%%%%%%%%%%%%%%%%%%%%%%%%%%%%%%%% 
Remind the reader that  the operator describing 
the correlation function contains the projector $\tY$:   
$$
Y(\lambda,\mu)
=P(\lambda) P(\mu)=
\stint\,dudvE_+(\lambda|v) E_+(\mu|u).
$$
In  the operator notations it can be written in the form
\be{projector}
\stint\,dudvE_+(\lambda|v) E_+(\mu|u)=
\EL\hat\sigma_-\ER,
\ee
where  the operator $\hat\sigma_-$ is equal to
$$
\hat\sigma_-=\left(
\begin{array}{cc}
0&0\\1&0
\end{array}
\right).
$$
(This is  the particular case of operator --- its discrete components are
constant functions of continuous  indices $u$ and $v$. The action 
of this operator on vectors is still given by \Eq{action}). 

Let us calculate  the derivative of 
$\det\left(\tilde I+\tV-\frac{\alpha}{2\pi}\tY\right)$ 
with respect to $\alpha$
$$
\begin{array}{c}{\left.\dis
\partial_\alpha\log\det\left(\widetilde I+
\tV-\frac{\alpha}{2\pi}\tY\right)\right|_{\alpha=0}}\\
\five\\ 
{\dis
=-\frac{1}{2\pi}\TR\left(\EL\hat\sigma_-\ER\circ\invop\right)
=-\frac{1}{2\pi}\TR\left(\EL\hat\sigma_-\FRl\right)}\\
\five\\ 
{\dis
=-\frac{1}{2\pi}\Tr\left(\hB\hat\sigma_-\right)
=-\frac{1}{2\pi}\stint B_{12}(u,v)\,dudv}.
\end{array}
$$
Here we denote by  the symbol ``$\TR$" the trace of 
the integral kernels. Hence,
\be{derivalpha}
\left.\partial_\alpha\det\left(\widetilde I+
\tV-\frac{\alpha}{2\pi}\tY\right)\right|_{\alpha=0}=
-\det\freop\stint \,\frac{dudv}{2\pi}B_{12}(u,v).
\ee
Due to  the definitions  \Eq{q}, \Eq{b} and identity 
\be{Ging}
G(x,t)=\frac{1}{2\pi}\stint\,dudvg(u,v),
\ee
we come to the equality
$$
G(x,t)-\frac{1}{2\pi}\stint B_{12}(u,v)\,dudv=
-\frac{1}{2\pi}\stint b_{12}(u,v)\,dudv.
$$
Hence  the correlation function is equal to
\be{correlator}
\langle\psi(0,0)\psi^\dagger(x,t)\rangle_T=
-\frac{e^{-iht}}{2\pi}
\brad\frac{\det\freop}{\det(\hat I-\frac1{2\pi}\widetilde K_T)}
\stint b_{12}(u,v)\,dudv\ketd.
\ee

The logarithmic derivatives of  the determinant of the operator
$\freop$ with respect to $x$ and $t$ are equal
$$
\begin{array}{c}{\dis
\partial_x\log\det\freop=\TR\left(
\partial_x\tV\circ\invop\right),}\\
\five\\
{\dis
\partial_t\log\det\freop=\TR\left(
\partial_t\tV\circ\invop\right).}
\end{array}
$$
Using \Eq{derivativexV} we have
$$
\begin{array}{c}{\dis
\TR\left(\partial_x\tV\circ\invop\right)
=-\TR\left(\EL\hs\ER\circ\invop\right)}\\
\five\\
{\dis
=-\qint\,d\lambda(\EL\hs\FRl)=-\Tr\biggl(\hB\cdot\hs\biggr),}
\end{array}
$$
and hence
\be{derivativexdet}
\partial_x\log\det\freop=-\Tr\biggl(\hB\cdot\hs\biggr).
\ee

In the same method one can compute the time derivative
$$
\begin{array}{c}{\dis
\TR\left(\partial_t\tV\circ\invop\right)}\\
\five\\
{\dis
=\TR\left(\EL(\hs(2\lambda+\mu-\lambda)+[\hg,\hs])\ER\circ\invop\right)}\\
\five\\
{\dis
=\qint\,d\lambda\biggl(2\lambda\EL\hs\FRl+
\EL[\hg,\hs]\FRl\biggr)}\\
\five\\
{\dis
-\qint\,d\lambda d\mu\biggl(\EL\hs\ER\FLm\FRls\biggr)}\\
\five\\
{\dis
=\Tr\biggl(2\hC\cdot\hs-\hB^2\cdot\hs+\hB\cdot[\hg,\hs]\biggr)
=\Tr\biggl(2(\hC+\hb\cdot\hg)\cdot\hs-\hb^2\cdot\hs\biggr).}
\end{array}
$$
Therefore
\be{derivativetdet}
\partial_t\log\det\freop
=\Tr\biggl(2(\hC+\hb\cdot\hg)\cdot\hs-\hb^2\cdot\hs\biggr).
\ee

The second logarithmic derivatives of  the determinant can 
be expressed in terms of matrix $\hb$ only. Indeed, 
using \Eq{ident1} and \Eq{ident2} we get
\be{dxdxdet}
\partial_x\partial_x\log\det\freop
=-\Tr\biggl([\hb,\hs]\cdot\hb\cdot\hs\biggr).
\ee
\be{dxdtdet}
\partial_x\partial_t\log\det\freop
=-\Tr\biggl(\partial_x\hb\cdot\hb\cdot\hs-
\hb\cdot\partial_x\hb\cdot\hs\biggr).
\ee
Thus, we have expressed the logarithmic derivatives of  the Fredholm
determinant in terms of the traces of operators $\hb$, $\hB$ and 
$\hC$.  It is worth mentioning that the second logarithmic derivatives 
depend on  the operator $\hb$ only.

Now let us turn back to the Lax representation.
The relations \Eq{derivativexF} and \Eq{derivativetF} 
$$
\begin{array}{c}
\partial_x\FRl=\widehat{\cal L}(\lambda)\FRl,\\
\five\\
\partial_t\FRl=\hat{\cal M}(\lambda)\FRl
\end{array}
$$
should be compatible. It means that
$$
\partial_t\widehat{\cal L}-\partial_x\hat{\cal M}
+[\widehat{\cal L},\hat{\cal M}]=0.
$$
Substituting \Eq{calL} and \Eq{calM1} into the last formula, we get
\be{compatible}
[\partial_t\hb,\hs]-\partial_x\partial_x\hb
+\biggl[[\hb,\hs],\partial_x\hb\biggr]=0.
\ee
{\sl Remark.}~~ 
More accurately the equation \Eq{compatible} is valid
if there exists a sequence of complex numbers $\lambda_n$ such that
  the corresponding sequence of vectors $|F^R(\lambda_n)\rangle$ 
generates a basis of the space ${\cal H}$. However, one can check 
that this condition is not necessary. Indeed, using identities for 
potentials from Appendix A (in particular formul\ae~\Eq{derivativexB} 
and \Eq{subderivtB}) it is possible to prove the equality 
\Eq{compatible} directly, without using the compatibility condition.

Thus, we have arrived at the following results. The second logarithmic
derivatives of the Fredholm determinant are presented in terms of  
  the operator $\hb$. On the other hand this operator satisfies  
the partial differential equation \Eq{compatible}. In turn this 
equation appears to be the compatibility condition of the auxiliary 
linear problem \Eq{derivativexF}, \Eq{derivativetF}.

%%%%%%%%%%%%%%%%%%%%%%%%%%%%%%%%%%%%%%%%%%%%%%%%%%%%%%
\section{Main results in components of operators $\hb$ and $\hC$}
%%%%%%%%%%%%%%%%%%%%%%%%%%%%%%%%%%%%%%%%%%%%%%%%%%%%%%
 In this Section we summarize the main results obtained in the
previous Sections. Till now we did not use the symmetry of the
kernel $V(\lambda,\mu)=V(\mu,\lambda)$ \Eq{kernelV}. Using this 
property and  the definitions \Eq{identification} one can get 
additional identity 
\be{adident} B_{11}(u,v)=-B_{22}(v,u).  
\ee 

Due to  the definitions \Eq{b} and \Eq{q} we have 
\be{bandB} 
\begin{array}{lcr}
{\dis
b_{ab}(u,v)=B_{ab}(u,v),}&\qquad&\hskip-1cm
\mbox{for all $a,b$ exept the case $a=1, b=2$},\\
\one &\one &\one\\
{\dis
b_{12}(u,v)=B_{12}(u,v)-g(u,v)}&&
\end{array}
\ee
It follows from  \Eq{derivativexB}, \Eq{trace} and  the definition 
\Eq{sigma} that 
\ba{components2}
\partial_xB_{11}(u,v)&=&
i\stint\,dwb_{12}(u,w)b_{21}(w,v),\\
\label{components3}
\partial_xB_{22}(u,v)&=&
-i\stint\,dwb_{21}(u,w)b_{12}(w,v).
\ea

Using identities \Eq{ident1}, \Eq{ident2}  and expressions 
for  the logarithmic derivatives of the determinant 
\Eq{derivativexdet}--\Eq{dxdtdet} we have

\ba{logdirx}
\partial_x\log\det\freop&=&
i\stint\,dub_{11}(u,u),\\
\label{logdirt}
\partial_t\log\det\freop&=&
i\stint\,dudv\Bigl(-b_{21}(u,v) g(v,u) \non
&+&i\stint\,du(C_{22}(u,u)-C_{11}(u,u))\Bigr),\\ 
\label{logdirxx}
\partial_x\partial_x\log\det\freop&=&
-\stint dudvb_{12}(u,v)b_{21}(v,u),\\
\label{logdirxt}
\partial_t\partial_x\log\det\freop&=&
i\stint dudv(\partial_x b_{12}(u,v)\cdot b_{21}(v,u)\non
&-&\partial_x b_{21}(u,v)\cdot b_{12}(v,u)).
\ea

The diagonal part of  the compatibility condition \Eq{compatible} is
equal to zero due to \Eq{components2} and \Eq{components3}. 
For the off-diagonal part of  the compatibility condition we find 
\ba{equat1}
-i\partial_tb_{12}(u,v)&=&-\partial_x^2b_{12}(u,v)\non
&&\hskip-2cm
+2\stint\,dw_1dw_2b_{12}(u,w_1)b_{21}(w_1,w_2)b_{12}(w_2,v),\\
\label{equat2}
i\partial_tb_{21}(u,v)&=&-\partial_x^2b_{21}(u,v)\non
&&\hskip-2cm
+2\stint\,dw_1dw_2b_{21}(u,w_1)b_{12}(w_1,w_2)b_{21}(w_2,v).
\ea
The system of partial differential equations \Eq{equat1} and 
\Eq{equat2} is  continuum generalization of  the classical
nonlinear Schr\"odinger equation. The second logarithmic derivatives 
of the Fredholm determinant \Eq{logdirxx} and \Eq{logdirxt} give us
the densities  of the first and the second integrals of motion of
 this equation. It means that the correlation function possess 
the properties of a $\tau$-function of the generalized nonlinear 
Schr\"odinger equation.

%%%%%%%%%%%%%%%%%%%%%%%%%%%%%%%%%%%%%%%%%%%%%
\section*{Summary}
%%%%%%%%%%%%%%%%%%%%%%%%%%%%%%%%%%%%%%%%%%%%%%%
The main goal of this paper was a description of the quantum 
correlation function in terms of solutions of the classical 
completely integrable  equation. We have constructed such 
differential equations---\Eq{equat1}, \Eq{equat2}, which, in fact, 
are continual generalization of nonlinear Schr\"odinger equation. The 
correlation function of local fields plays role of  the  
$\tau$-function of these equations. In particular the second 
logarithmic derivatives of the Fredholm determinant \Eq{logdirxx} and 
\Eq{logdirxt} give us the densities  of the first and the second 
integrals of motion.  In a forthcoming publication we shall 
formulate the Riemann--Hilbert problem for  the differential equation 
obtained in this paper. This permit us to evaluate the long distance 
asymptotic.

%%%%%%%%%%%%%%%%%%%%%%%%%%%%%%%%%%%%%%%%%%%%%
\section*{Acknowledgments}
%%%%%%%%%%%%%%%%%%%%%%%%%%%%%%%%%%%%%%%%%%%%%%%
We wish to thank Professor T. Inami for usefull discussions.
This work is partly supported by the National Science Foundation (NSF)
under Grant No. PHY-9605226, the Japan Society for
the Promotion of Science, the Russian Foundation of Basic Research under
Grant No. 96-01-00344 and INTAS-01-166-ext.

\appendix
%%%%%%%%%%%%%%%%%%%%%%%%%%%%%%%%%%%%%%%%%%%%%%%%%%%%%%
\section{Identities for potentials}
%%%%%%%%%%%%%%%%%%%%%%%%%%%%%%%%%%%%%%%%%%%%%%%%%%%%%
Here we establish some identities between matrix elements of
potentials $\hB$ (or $\hb$) and $\hC$. We also introduce
 the operator $\hD$:
\be{hD}
\hD=\qint\lambda^2\FRl\EL\,d\lambda.
\ee

Let us calculate the derivative of the operator   $\hB$ with 
respect to $x$.  We shall use formul\ae\ \Eq{derivativexE}, \Eq{L}, 
\Eq{derivativexF}  and \Eq{calL}. We have
\ba{subderivxB} &&{\dis 
\partial_x\hB=\qint\,d\lambda\left(
\widehat{\cal L}(\lambda)\FRl\EL
-\FRl\EL\widehat L(\lambda)\right)}\non
\one & \one & \one \non
&&={\dis
\qint\,d\lambda\left((\lambda\hs+[\hb,\hs])
\FRl\EL-\FRl\EL (\lambda\hs+[\hg,\hs]\right)}\non
\one & \one & \one \non
&&={\dis
[\hs,\hC]+[\hb,\hs]\cdot\hB-\hB\cdot[\hg,\hs].}
\ea 
Therefore
\be{derivativexB}
\partial_x\hB=[\hs,\hC]+[\hb,\hs]\cdot\hB-\hB\cdot[\hg,\hs].
\ee
Taking the trace of this equality we have
\be{trace}
\Tr\biggl(\partial_x\hB\biggr)=0.
\ee
Here we used definition $\hb=\hB+\hg$.
Multiplying \Eq{derivativexB} by $\hs$ and taking the trace 
we have
\be{ident1}
\Tr\biggl(\partial_x\hB\cdot\hs\biggr) =
\Tr\biggl([\hb,\hs]\cdot\hb\cdot\hs\biggr).
\ee
Here we used the property
$$
\hg^2=(\hg\hs)^2=0,\qquad\mbox{see \Eq{q}}.
$$
Using the same method one can calculate the derivative of  the  
operator $\hB$ with respect to  $t$. To do it we need formul\ae\ 
\Eq{derivativetE}, \Eq{M}, \Eq{derivativetF}  and \Eq{calM}. We have 
\ba{subderivtB} &&{\dis 
\partial_t\hB=\qint\,d\lambda\left(
\hat{\cal M}(\lambda)\FRl\EL-\FRl\EL\hat M(\lambda)\right)}\non
\one & \one & \one \non
&&={\dis
-[\hs,\hD]-[\hb,\hs]\cdot\hC
+\hC\cdot[\hg,\hs]+\partial_x\hb\cdot\hB
-\hB\cdot\partial_x\hg.}
\ea 
Again multiplying by $\hs$ and taking the trace we have
$$
\Tr\biggl(\partial_t\hB\cdot\hs\biggr)=
\Tr\biggl([\hs,\hC]\cdot\hg\cdot\hs
-[\hs,\hC]\cdot\hs\cdot\hb+\partial_x\hb\cdot\hB\cdot\hs
-\hB\cdot\partial_x\hg\cdot\hs\biggr).
$$
(Here we used cyclic permutation under the sign ``$\Tr$"). 
Substituting $[\hs,\hC]$ from \Eq{derivativexB} we get after simple 
algebra 
\be{ident2} \Tr\biggl(\partial_t\hB\cdot\hs\biggr)= 
\Tr\biggl(\partial_x\hb\cdot\hb\cdot\hs-
\hb\cdot\partial_x\hb\cdot\hs\biggr).
\ee

These identities were used for calculation of the second logarithmic
derivatives of the Fredholm determinant in  Section 6. One can also 
use the method described above in order to prove directly  the  
equation \Eq{compatible} 
$$ 
[\partial_t\hb,\hs]-\partial_x\partial_x\hb
+\biggl[[\hb,\hs],\partial_x\hb\biggr]=0.
$$
%%%%%%%%%%%%%%%%%%%%%%%%%%%%%%%%%%%%%%%%%%%%%%%%%%%%%%%
\section{Quantum equation as continual differential equation }
%%%%%%%%%%%%%%%%%%%%%%%%%%%%%%%%%%%%%%%%%%%%%%%%%%%%%%%%
 Consider the quantum nonlinear Schr\"odinger equation
\be{quantNLS}
i\partial_t\psi=-\partial_x^2\psi+2c\psi^\dagger\psi\psi,
\ee
where $\psi(x,t)$ is annihilation operator in the bosonic Fock space. 
Let us take a matrix element of this equation between the ground 
state at finite density $|\Omega\rangle$ and state with one hole
$\langle h_0|$. We shall have in mind that the momentum of the hole
$k_0$ is fixed.
\be{contNLS}
i\partial_t\langle h_0|\psi|\Omega\rangle
=-\partial_x^2\langle h_0|\psi|\Omega\rangle
+2c\langle h_0|\psi^\dagger|h_0,h_1\rangle
\langle h_1,h_0|\psi|h_2\rangle\langle h_2|\psi|\Omega\rangle.
\ee
Here $h_2$ is a hole with momentum $k_2$ and $h_1$ is a hole with
momentum $k_1$. In the last term in the r.h.s. of \Eq{contNLS} we
have to integrate with respect to $k_1$ and $k_2$. This shows that 
quantum nonlinear Schr\"odinger equation is equivalent to the 
classical continual differential equation. The same type of equations
describes quantum correlation functions (see \Eq{equat1}, 
\Eq{equat2}).

%%%%%%%%%%%%%%%%%%%%%%%%%%%%%%%%%%%%%%%%%%%%%%%%%%%%%%% 
\section{Free fermion limit }
%%%%%%%%%%%%%%%%%%%%%%%%%%%%%%%%%%%%%%%%%%%%%%%%%%%%%%%%

In the free fermion limit $c\to\infty$ function $\EM\lambda$  
\Eq{eminus} does not depend on continuous variable $u$. Indeed, in this
limit we have $h(\lambda,\mu)=(\lambda-\mu+ic)/ic\to 1$, hence all 
commutators \Eq{commutators} are equal to zero. Thus, one can put 
all dual fields equal to zero in expressions for functions $E_\pm$ % 
\ba{eplusfree}
\EP\lambda&=&{\dis\frac{\sqrt{\theta(\lambda)}}{2\pi}
\left(\frac{1}{u-\lambda+i0}
+\frac{1}{u-\lambda-i0}\right)
e^{\tau(u)-\frac12\tau(\lambda)},}\\
\one &\one &\one \non
\label{eminusfree}
\EM\lambda&=&{\dis\frac{1}{\pi}
e^{-\frac12\tau(\lambda)}\sqrt{\theta(\lambda)}}.
\ea

Hence matrix elements of
 the  operator $\hB$ possess the following properties: 
$B_{21}(u,v)$ does not depend on $u$ and $v$; $B_{11}(u,v)$ depends 
only on the first argument $u$; $B_{22}(u,v)$ depends only on the 
second argument $v$; $B_{12}(u,v)$ depends on both arguments $u$ and 
$v$. Analogous properties have matrix elements of 
 the operator $\hC$.

One can make the replacement
$$
\begin{array}{c}
B_{21}(u,v)\to B_{21},\\         
\stint\,duB_{11}(u,v)\to B_{11},\\         
\stint\,dvB_{22}(u,v)\to B_{22},\\         
\stint\,dudvB_{12}(u,v)\to B_{12},
\end{array}
\qquad
\begin{array}{c}
C_{21}(u,v)\to C_{21},\\         
\stint\,duC_{11}(u,v)\to C_{11},\\         
\stint\,dvC_{22}(u,v)\to C_{22},\\         
\stint\,dudvC_{12}(u,v)\to C_{12}.
\end{array}
$$
Here functions $B_{ab}$ and $C_{ab}$ are scalar potentials.
In terms of these potentials one can rewrite  the correlation function, 
the logarithmic derivatives of determinant and the partial 
differential equations in the form 
$$ 
\langle\psi(0,0)\psi^\dagger(x,t)\rangle_T=
-\frac{e^{-iht}}{2\pi}b_{12}\det\freop.
$$
\vskip0.5cm
\centerline{The logarithmic derivatives of  the determinant}
$$
\begin{array}{rcl}
\partial_x\log\det\freop&=&
ib_{11},\\
\one&\one&\one\\
\partial_t\log\det\freop&=&
i(-C_{11}+C_{22}-b_{21}G),\\ 
\one&\one&\one\\
\partial_x\partial_x\log\det\freop&=&
-b_{12}b_{21},\\
\one&\one&\one\\
\partial_t\partial_x\log\det\freop&=&
i(\partial_x b_{12}\cdot b_{21}
-\partial_x b_{21}\cdot b_{12}).
\end{array}
$$
\vskip0.5cm
\centerline{ The differential equations.}
(One should integrate \Eq{equat1} with respect to $u$ and $v$)
$$
\begin{array}{rcl}
-i\partial_tb_{12}&=&-\partial_x^2b_{12}+2b_{12}^2b_{21},\\
\one&\one&\one\\
i\partial_tb_{21}&=&-\partial_x^2b_{21}+2b_{21}^2b_{12}.
\end{array}
$$
Here $b_{ab}=B_{ab}$  for all $a$ and $b$ except $a=1,\quad b=2$; 
$b_{12}=B_{12}-G$.  
These results reproduce the results of \cite{IIKS} up to
notations and rescaling  the distance $x$ and the time $t$.

\end{document}